\documentclass[final,twocolumn]{revtex4} 

\usepackage{hyperref}
\usepackage{epsfig,graphicx}
\usepackage{dcolumn}   
\usepackage{amssymb}
\usepackage{amsfonts}
\usepackage{bm}
\usepackage{color}

\def\fun#1#2{\lower3.6pt\vbox{\baselineskip0pt\lineskip.9pt
\ialign{$\mathsurround=0pt#1\hfil ##\hfil$\crcr#2\crcr\sim\crcr}}}

\newcommand{\beq}{\begin{eqnarray}}
\newcommand{\eeq}{\end{eqnarray}}
\newcommand{\be}{\begin{equation}}
\newcommand{\ee}{\end{equation}}

\def\fun#1#2{\lower3.6pt\vbox{\baselineskip0pt\lineskip.9pt
\ialign{$\mathsurround=0pt#1\hfil ##\hfil$\crcr#2\crcr\sim\crcr}}}

\newcommand{{\SD}}{\rm SD}

\newcommand{\vesig}{\mbox{\boldmath${\rm \sigma}$}}

\newcommand{\vep}{\mbox{\boldmath${\rm p}$}}
\newcommand{\veq}{\mbox{\boldmath${\rm q}$}}

\newcommand{\lan}{\langle}
\newcommand{\ran}{\rangle}

\begin{document}

\title{Dynamical origin and the pole structure of  $X(3872)$}

\author{\firstname{I.V.}~\surname{Danilkin}}
\email{danilkin@itep.ru} \affiliation{Gesellschaft fur
Schwerionenforschung (GSI), Planck Str. 1, 64291 Darmstadt,
Germany}\affiliation{Institute of Theoretical and Experimental
Physics, Moscow, Russia}

\author{\firstname{Yu.A.}~\surname{Simonov}}
\email{simonov@itep.ru} \affiliation{Institute of Theoretical and
Experimental Physics, Moscow, Russia}

\begin{abstract}
The dynamical mechanism of channel coupling with the decay
channels is applied to the  case of coupled charmonium - $DD^*$
states with $J^{PC}=1^{++}$. A pole analysis is done and the
$DD^*$ production cross section is calculated  in qualitative
agreement with experiment. The sharp peak at the $D_0D^*_0$
threshold and flat background are shown to be due to Breit-Wigner
resonance, shifted by channel coupling from the original position
of 3954 MeV for the $2^3P_1$, $Q\bar Q$ state. A similar analysis,
applied to the $n=2$, $^3P_2$, $~^1P_1$, $^3P_0$, allows us to
associate the first one with the observed $Z(3930)$ $J=2$ and
explains the destiny of $~^3P_0$.

\end{abstract}

\maketitle


The resonance $X(3872)$ found in \cite{Choi:2003ue} and confirmed
and further studied in
\cite{Acosta:2003zx,Abazov:2004kp,Aubert:2004ns,Barnes:2003vb}
(see \cite{Pakhlova:2008di} for review) is still a mysterious
phenomenon. The measured quantum numbers of X(3872)
\cite{Acosta:2003zx} suggest that this is a $1^{++}$ state. One
can list several properties of this resonance which are difficult
to explain. (1) The width of the peak at  $3872$ MeV is zero
within experimental energy resolution. (2) The peak is exactly at
the $D_0D_0^*$ threshold ($3871.2$ MeV) and not at a little higher
$D_+D_-^*$ threshold ($3879$ MeV); however, isospin conservation
predicts that both thresholds should enter with the same weight.
(3) The single-channel theory \cite{Eichten_Badalian} predicts a
standard $2^{\,3}P_1$ level of the $Q\bar Q$ system around 3950
MeV; however, among  the structures observed by Belle in this
region, $X(3940), Y(3940), Z(3930)$, there seems to be no examples
suggesting the $1^{++}$ identification \cite{Pakhlova:2008di}. (4)
Among the four members of the $2^{3,1} P_J$ multiplet, only one
with $J=2$ can be associated with $Z(3930)$, which was observed as
a regular resonance,  $X(3872)$ looks like a sharp cusp, and two
others are not seen in this region. It is important to explain
this very different behavior.

On the theoretical side there are models based on the $D_0D_0^*$
molecular picture of $X(3872)$
\cite{Tornqvist:2004qy_Close:2003sg,Voloshin:2003nt,Swanson:2003tb,Braaten:2005ai}
and the tetraquark system \cite{Maiani_Valcarce_Vijande}; see
\cite{Godfrey:2009qe} for a review. However, one cannot get a
simultaneous explanation of points (1)-(4) from these models, and
we develop here an alternative approach. It is a purpose of this
Letter to exploit a realistic dynamical mechanism, constructed in
\cite{Danilkin:2009hr}, which can explain all four points. Below
we shall Briefly explain the mechanism of channel coupling (CC)
with the decay channels \cite{Danilkin:2009hr}. This method allows
us to calculate not only position of poles in the CC system, like
$Q\bar Q$ and $Q\bar{q}$, $\bar{Q}q$, but also scattering
amplitudes and production cross sections. We demonstrate that, in
the $2^{\,3}P_1$ CC system coupled by the $S$-wave decays, two
poles originating from complex conjugate Breit-Wigner resonances
of $Q\bar Q$ system are shifted by CC to the final position, with
one pole yielding a narrow cusp at one of thresholds, and another
a shifted flat background. We show that when the $CC$ coupling
increases, the latter pole yields a very broad bump, and at the
same time the weak threshold cusp  at the higher threshold
$D_+D_-^*$ goes over into a sharp peak at the lower one
$D_0D_0^*$. In this way the same value of the coupling constant
$\gamma$ (well within the accuracy limits of the universal
constant fitted to different charmonium and bottomonium states in
\cite{Simonov:2007bm}) produces the visible effects, compatible
with the properties (1)-(4) mentioned above. To produce this
effect as in the $2^{\,3}P_1$ state, the original single-channel
pole should be above and in the attraction region of the
threshold. For the poles originally below threshold, CC shifts
poles down, as it happens with the $2^{\,3}P_2$ pole. The same
approach allows us to explain the situation with two other poles,
$2^{\,3}P_0$ and $2^{\,1}P_1$, as will be discussed below.

Resonances in coupled channels can exist for 3 different reasons
\cite{Badalian:1981xj}: (a) due to bound states in the $Q\bar Q$
channel, which are shifted by CC, (b) due to poles in the $(Q\bar
q) (\bar Qq)$ channel, shifted by CC; (c) due to strong CC alone
(even if no  interaction exists in decoupled channel). The most
striking feature of the CC resonance is that it approaches the
threshold at increasing coupling and typically looks like a
pronounced cusp at the threshold of small width. In the realistic
physical problem several of these reasons can be present at the
same time: e.g., the bare state in the single-channel charmonium
can be shifted by strong CC exactly to the threshold. This
situation will be discussed below and is characteristic for the
single-channel pole above  the $S$-wave threshold, where one meets
at the same time with the $Q\bar Q$, and the strong CC
interaction, which shifts the bare charmonium pole exactly to the
threshold position and produces a sharp peak. We shall
quantitatively describe this situation in the CC formalism of
\cite{Danilkin:2009hr}, where the only changeable parameter is the
channel coupling constant $\gamma$ being varied around the
standard value.

The basis of the  CC theory developed in
\cite{Danilkin:2009hr,Simonov:2007bm} can be shortly formulated in
three relations: a) The effective string decay Lagrangian of the
$^3P_0$ type for the decay $Q\bar Q\to (Q\bar{q})(\bar{Q}q)$
\begin{equation}\label{Eq.1}
\mathcal{L}_{sd} =\int \bar \psi_q M_\omega \psi_q  \,d^4x
\end{equation}
with $M_\omega$ tested in charmonium and bottomonium decays
\cite{Simonov:2007bm}, $ M_\omega\approx 0.8$ GeV where light
quark bispinors are treated in the limit of large $m_c$ mass as
solutions of Dirac equations, and this allows us to go over to the
reduced $(2\times 2)$ form of the decay matrix element,
$\mathcal{L}_{sd} \to \gamma \int iv_c\, \vesig \vep\, v\,  d^4
x$, $\gamma =\frac{M_\omega}{\lan m_q +U-V+\varepsilon_0\ran }
\approx 1.4$ (with realistic averages of scalar $U=\sigma r$ and
vector $V=-\frac34 \frac{\alpha_s}{r}$ potentials). b) The decay
matrix element of the state $n_1$ of heavy quarkonium $Q\bar Q$ to
the states $n_2$ and $n_3$ of heavy-light mesons $Q\bar q$, $\bar
Qq$,
\begin{eqnarray}
\nonumber J_{n_1n_2n_3}(\vep)=&&\frac{\gamma}{\sqrt{N_c}} \int
\frac{d^{\,3} q}{(2\pi)^3}\,\bar y^{red}_{123} (\vep, \veq)\,
\Psi^{+(n_1)}_{Q\bar Q} (c\vep+\veq)\\ && \times\,\psi_{Q\bar
q}^{(n_2)} (\veq)\, \psi^{(n_3)}_{\bar Q q} (\veq). \label{Eq.2}
\end{eqnarray}
Here all wave functions $(\Psi_{Q\bar Q},\, \psi_{Q\bar q},\,
\psi_{\bar Q q})$ refer to the radial parts of the corresponding
wave functions, while $\bar y^{red}_{123}$ comprises the decay
vertex of $\mathcal{L}_{sd}$ and all spin-angular parts of mesons
involved; the list of $\bar y_{123}^{red}$ for the 6 lowest states
is given in Table VII of \cite{Danilkin:2009hr}. In Eq.\ref{Eq.2}
$c=\frac{\omega_Q}{\omega_q+\omega_Q}$, where the averaged kinetic
energies of heavy and light quarks in D meson
$\omega_{q}\simeq0.55$ GeV, $\omega_{Q}\simeq1.5$ GeV are taken
from \cite{Badalian:2007km}. c) The CC interaction ``potential''
$V^{CC}_{n_2n_3}$ between $Q\bar q$ and $\bar Qq$ mesons due to
intermediate states of bound $Q\bar Q$ system,
\begin{equation}\label{Eq.3}
V^{CC}_{n_2n_3} (\vep, \vep', E) = \sum_n \frac{J^+_{nn_2n_3}
(\vep) J_{nn_2n_3}(\vep')}{E-E_n}
\end{equation}
and the final $(Q\bar q)(\bar Qq)$ Hamiltonian looks like $H=H_0+
V^{CC}$, where $H_0$ may contain the direct $Q\bar q$ and $\bar
Qq$ interaction, which is $\mathcal{O}(1/N_c)$ and we disregard it
in what follows. The equation for the pole position is
\cite{Danilkin:2009hr}
\begin{equation}\label{Eq.4}
\det[E- \hat E- {\hat w(E)}] =0, ~~ (\hat E)_{mn} = E_n\,
\delta_{mn}
\end{equation}
where $E_n$ is the mass of the bare states and $w_{nm} (E)$ is
\begin{equation}\label{Eq.7}
w_{nm} (E) = \int \frac{d^{\,3}\vep}{(2\pi)^3} \sum_{n_2, n_3}
\frac{J_{nn_2n_3}(\vep) J^+_{mn_2n_3}
(\vep)}{E-E_{n_2n_3}(\vep)}\,,
\end{equation}
with $E_{n_2n_3}(\vep)=E_{n_2}+E_{n_3}.$ We shall be interested
below in the pole positions, i.e., solutions of (\ref{Eq.4}), and
the most important for comparison with experiment, is the
production cross section
\begin{eqnarray}\label{Eq.8}
\sigma_{prod} = \sum_n \Phi_n(E)\frac{\mathrm{Im}\, w_{nn}
(E)}{|E-E_n-w_{nn} (E)|^2},
\end{eqnarray}
where $\Phi_n(E)$
is a weakly changing function of E. In
Figs.\ref{Production_2P_two} and Fig.\ref{fig.1P1} this factor is
omitted.

We first  apply this method to the case of the $1^{++}$ state of
$Q\bar Q$, and we confine ourselves to one state $2^3P_1$ in the
$Q\bar Q$ system so that Eq. (\ref{Eq.4}) reduces to
\begin{equation}\label{Eq.9}
E-E_n= w_{nn} (E)
\end{equation}
and $w_{nn}$ is given in (\ref{Eq.7}), where
$J_{nn_2n_3}(\mathbf{p})$ should be calculated with the wave
functions of $D$ and $D^*$ mesons (we disregard here the possible
difference in wave functions  of $D_0$ and $D_\pm$, and of $D_0^*$
and $D_\pm^*$, and we also disregard all other states $n_2, n_3$
beyond $DD^*$). The wave functions of all states involved have
been calculated  by Badalian et al. \cite{Eichten_Badalian}, using
the relativistic string Hamiltonian \cite{Dubin:1994vn} derived in
the framework of the field correlator method \cite{Dosch:1987sk}.
Here only universal input is used: current quark masses $m_q$,
string tension $\sigma$ and strong coupling $\alpha_s$. These
realistic w.f. $(\Psi_{Q\bar Q},\, \psi_{Q\bar q},\, \psi_{\bar Q
q})$ have been fitted by a series of oscillator w.f. and in this
way both $J_{nn_2n_3}(\vep)$ and $w_{nn} (E)$ were numerically
obtained.

\begin{figure}[t]
  \includegraphics[angle=0,width=0.5\textwidth]{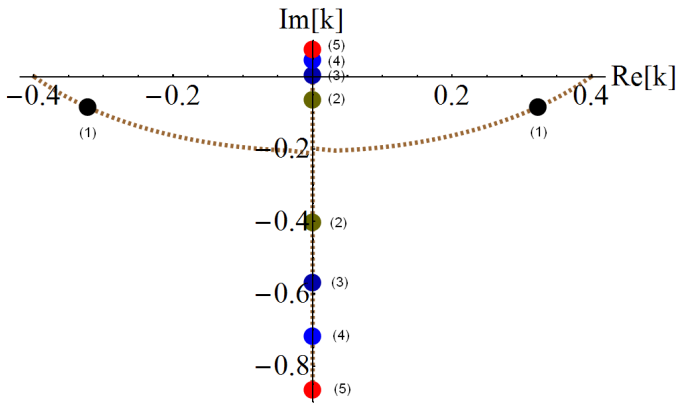}
  \caption{Motion of poles $k_+$ (starting on right-hand side) and $k_-$
 (starting on left-hand side) in the $k$ plane (in units GeV) with growing coupling
 parameter $\gamma$. Numbers at field circles on curves
 correspond to numbers on curves in Fig. \ref{Production_2P_two}a.}
  \label{fig.poles}
\end{figure}

\begin{figure}[t]
\begin{minipage}[h]{1.0\linewidth}
  \center{\includegraphics[angle=270,width=1.0\textwidth]{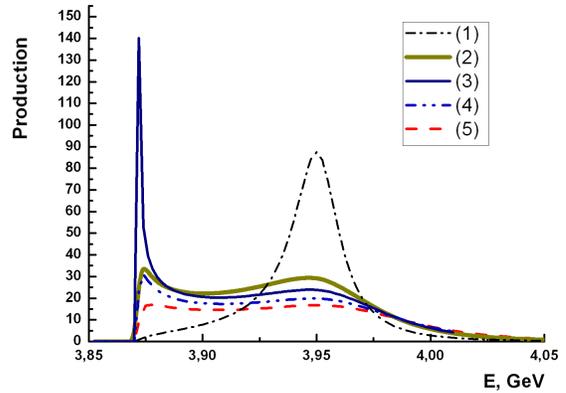}
  (a) One threshold, $E_{th}(D_0D^*_0)$=3.872 GeV.}
  \end{minipage}
  \begin{minipage}[h]{1.0\linewidth}
  \center{\includegraphics[angle=0,width=1.0\textwidth]{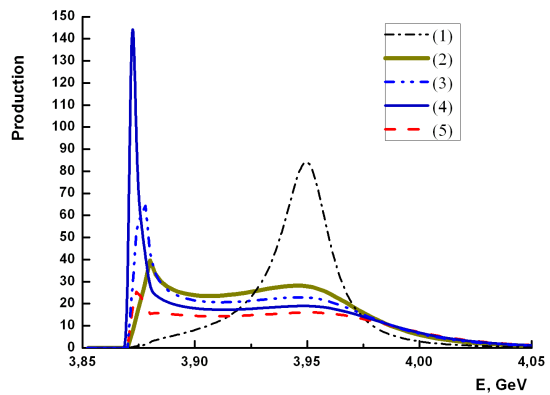}
  (b) Two thresholds, $E_{th}(D_0D^*_0;D_+D_-^*)$=3.872; 3.879 GeV.}
  \end{minipage}
\caption{Production [see Eq. \ref{Eq.8}] (in units GeV$^{-1}$) for
  $1^{++}$ state with different values of channel coupling parameter [(1) $\gamma=0.6$, (2) $\gamma=1.0$, (3) $\gamma=1.1$, (4)
$\gamma=1.2$, (5) $\gamma=1.3$].  For small values of channel
coupling parameter $\gamma$ [curve (1)] one can see a good
Breit-Wigner shape,  which corresponds to the shifted $2^{\,3}P_1$
state, while for larger $\gamma$ [curves (3) and (4)] there is a
broadening of higher resonance together with steep rise near the
threshold $E_{th}(D_0D^*_0)=3.872$ GeV.
}
  \label{Production_2P_two}
\end{figure}

\begin{figure}[t]
  \includegraphics[angle=0,width=0.5\textwidth]{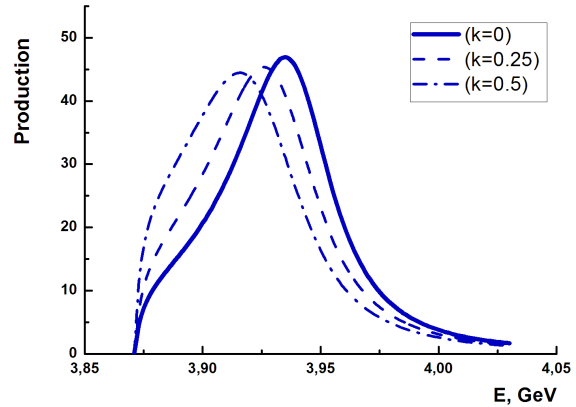}
  \caption{Production (see Eq. \ref{Eq.8}) [in units GeV$^{-1}$]
  for the  $1^{+-}$ state in $DD^*$ channel for channel coupling parameter $\gamma=1.1$ and with
  different values of closed $D^*D^*$ channel suppression coefficient, $\kappa=0;0.25;0.5$.}
  \label{fig.1P1}
\end{figure}

To understand  the nature of singularities in the energy plane,
which produce the cusp at the $D_0D_0^*$ threshold, we find
solutions of (\ref{Eq.9}). We separate from $w(E)$ the square root
singularity,  while the rest is  a slowly varying function, which
we approximate by $w(E_{th})$
\begin{equation}\label{Eq.11}
w(E) \cong  w(E_{th}) -\frac{i\tilde{M}}{2\pi}\, k\,|J(0)|^2
\end{equation}
where $E=E_{th}+k^2/2\tilde{M}$ and
$\tilde{M}=\frac{M_DM_{D^*}}{M_D+M_{D^*}}$ is the reduced mass.
Note that $w(E_{th})<0$, and the two pole solutions of
Eq.(\ref{Eq.9}) are
\begin{equation}\label{Eq.13}
k_{\pm} = -\frac{ia}{2} \pm \sqrt{-\frac{a^2}{4} +b}.
\end{equation}
with $a = \tilde M^2 |J(0)|^2/\pi$ and $b = 2 \tilde M\,
[w(E_{th}) +(E_n-E_{th})]$. Starting with small coupling, one has
two Breit-Wigner poles. With increasing $\gamma$ the square root
vanishes at $\gamma=\bar \gamma$ and the two poles collide; for
$\gamma>\bar \gamma$, both poles move apart along imaginary $k$
axis as shown in Fig.\ref{fig.poles}; and at some
$\gamma=\gamma^*,  ~ \gamma^*=\frac{E_n-E_{th}}{|w(E_{th})|},$ the
pole $k_+$ passes zero, providing the sharp peak at the threshold.

The analysis can be extended to the case of two thresholds
$E_{th}^{(1)}$ and $E^{(2)}_{th}$ with the resulting equation
\begin{eqnarray}\label{Eq.16}
\nonumber \left(k^2_1-b+ ik_1\frac{a}{2}\right)^2  +
(k^2_1-\Delta)\left(\frac{a}{2}\right)^2 =0,\end{eqnarray}
where $\Delta\equiv 2\tilde M (E^{(2)}_{th} - E_{th}^{(1)})$. For
small $\Delta$ the analysis goes as before, and the only
difference in the $k_1$ plane is the appearance of  the cut
connecting points $k_1=\pm \sqrt{\Delta}$, which denote access to
the second sheet of $k_2$. As before, the trajectory of the
highest pole $(k^{(+)}_1)$ passes through the origin, leading to a
sharp cusp at $E^{(1)}_{th}$. We have found that the pole never
passes through the point $E_{th}^{(2)}$, implying that the peak at
$E^{(2)}_{th}$ is never so high, as  at $E^{(1)}_{th}$, compare
curves (2) and (4) in Fig.\ref{Production_2P_two}b. These curves
correspond to the situations when the pole is closest to
$E_{th}^{(2)}$ and when the pole passes the origin, respectively.

Summarizing this analysis, we have found the pole structure behind
the phenomenon of $X(3872)$, and we may assert that the sharp peak
at 3872 is due to the  pole $k^{(+)}_1$ very close to the
$D_0D^*_0$ threshold, which originally was a genuine Breit-Winger
pole, and the flat background contains the far virtual pole
$k^{(-)}_1$, originally the complex conjugated Breit-Wigner pole
generated by the same  charmonium $2^3P_1$ state at $\sim 3950$
MeV, and shifted to the final position $k^{(-)}_1$ by CC. We
stress that the necessary condition for the threshold cusp of the
type of $X(3872)$ is $|E_n-E_{th}| \approx |w(E_{th})|$, which
means that the strength of CC should be as large as the distance
of the original $Q\bar Q$ bound state from the threshold. In other
words, the $Q\bar Q$ pole should be ``within reach'' of CC
interaction. One can also see from (\ref{Eq.13}) that for
$E_n<E_{th}$ the radicand is negative, and for the growing
$\gamma$ the pole $k_+$ moves up, farther from threshold, so that
for moderate $\gamma$ both poles are far from threshold.

In the case of two distinct thresholds $w(E)$ contains two
isotopically equivalent thresholds, which we take into account
with equal weights. To compare with experiment, we have used the
$Q\bar Q$ production cross section (\ref{Eq.8}), where $Q\bar Q$
(in our case $c\bar c$) are produced in some primary reaction,
e.g. in $e^+e^-$ double charmonium production or from $B\to KX$
and then the $Q\bar Q\to (Q\bar{q})(\bar{Q}q)$ transition takes
place. The resulting form of the cross section is shown in
Fig.\ref{Production_2P_two}b for five different values of
$\gamma$. One can see from Fig.\ref{Production_2P_two}(b) that for
weak coupling [curve (1)] only the single-channel charmonium state
$E_{n} (2^3P_1)$ is seen with $\Gamma\sim 35$ MeV, and the next
two curves display a cusp at $E^{(2)}_{th}$ and an almost
disappeared $Q\bar Q$ resonance, while curve (\ref{Eq.4}) clearly
signals a strong cusp at $E^{(1)}_{th}$ and no other features. At
even stronger CC [curve (5)] the CC pole goes away from thresholds
and the whole picture flattens. Thus we see that the experimental
situation is well reproduced by curve (4). The positions of both
poles changing with $\gamma$ are marked in Fig.\ref{fig.poles}.
One can easily see how the pole $k_+$ produces the sharp cusp in
position (3) for one threshold treatment, corresponding to curve
(3) in the production cross section in
Fig.\ref{Production_2P_two}(a).

Having found the mechanism, producing the peak at the $D_0D_0^*$
threshold, one may wonder what happens with other states of the
$n=2 ~^3P_J$ family,  $J=0,2$. To  this end one should first
estimate the position of  bare poles $E_n$ (see Table
\ref{tab.2}). We use the results of Badalian et al.
\cite{Eichten_Badalian} with the slightly modified spin-orbit
interaction.

\begin{table}[t]
\caption{\label{tab.2} Hadronic shift (MeV) of charmonium
$2^{3,1}P_J$ bare states $E_n$ for different channels. The bare
positions were taken from Badalian et al. \cite{Eichten_Badalian},
$\delta$ is the total shift and $\kappa$ is the closed channel
($D^*D^*$) suppression coefficient.}
\begin{center}
\begin{tabular}{c|c|c|cclc|c|cc}
\hline\hline
        State  & $J^{PC}$ &$E_n$ &     \multicolumn{5}{c|}{Shifts $(\gamma=1.1)$}                              &~\quad    $E$ ~\quad & Exp.\\
               &          &      &   $DD$  & $DD^*$  & \multicolumn{2}{c}{$D^*D^*$} & $\delta$                          &      & \\
\hline
              &  &        & -  &-                    & $\kappa=0$ &0     & 0   &     3.969  & \\
    $2 ^3P_2$ & $2^{++}$  & 3969   & -  &-                    & $\kappa=0.25$& -14 & -14  &     3.955 & Z(3930)\\
              &  &        & -  &-                    & $\kappa=0.5$&  -27 & -27  &     3.942 & \\
\hline
    $2 ^3P_1$ & $1^{++}$  & 3954   & \multicolumn{4}{c}{to threshold}     &     &3.872 & X(3872)\\
\hline
              &  &        & -2  &-                    & $\kappa=0$& 0     & -2  &      3.916  & \\
    $2 ^3P_0$ & $0^{++}$  & 3918   & -2  &-                    & $\kappa=0.25$& -33 & -35  &     3.883 & -\\
              &  &        & -2  &-                    & $\kappa=0.5$& -66  & -68  &     3.850 & \\
\hline
              &  &        & -  &-25                   & $\kappa=0$& 0     & -25  &     3.934 & \\
    $2 ^1P_1$ & $1^{+-}$  & 3959   & -  &-29                   & $\kappa=0.25$& -7  & -36  &     3.927 & -\\
              &  &        & -  &-32                   & $\kappa=0.5$& -14  & -46  &     3.918 & \\
\hline\hline
\end{tabular}
\end{center}
\end{table}

We take now the $2^3P_2$ bare state, which is mostly connected
with the $D^*D^*$ channel, while the $DD$ state is in the $D$ wave
and can be neglected. One can see, that $E_n(2^3P_2)<
E_{th}(D^*D^*)$, and $w(E)$ in (\ref{Eq.7}) is real and negative.
Hence, with increasing coupling the pole is shifted down, away
from the threshold. This is shown in Table \ref{tab.2}. In terms
of our previous analysis, using Eq. (\ref{Eq.13}) with $b<0$, one
can see that the pole is on the imaginary axis. Moreover, one can
estimate that $\sqrt{|b|}\gg\frac{a}{2}$ and the
near-the-threshold approximation (\ref{Eq.11}) is not applicable;
one can better use the original Eq. (\ref{Eq.9}), which yields a
shift of $|w(E_n)|\sim 55$ MeV. At this point one should take into
account the necessity of renormalizing the contributions of higher
closed thresholds  (which otherwise produce unacceptable shifts,
see \cite{Kalashnikova:2005ui,Danilkin:2009hr,Geiger:1992va}).
Therefore we introduce the coefficient $\kappa$, which multiplies
in (\ref{Eq.3}) the contribution of the closed channel $D^*D^*$.
We estimate $\kappa$ in an approximate range $0\leq\kappa\leq 0.5$
because it gives the most sensible results for the mass shift of
$J/\psi$. However $\kappa$ may depend on quantum state and bare
energy. The resulting position is near the experimentally found
\cite{Pakhlova:2008di} $Z(3930)$ peak, as seen in Table
\ref{tab.2}.

A similar  situation occurs for the $2^3P_0$ state, where the
lower threshold is far below, $|E_n(2^3P_0) - E_{th}(DD)|\sim 200$
MeV, while the higher threshold is more distant, than in the case
of the $2^3P_2$ state. Our calculation for the shift of the
$2^3P_0$ state yields a large value $\Delta E\sim 68$ MeV for
$\kappa =0.5$, see Table \ref{tab.2}, so the final position is
$3.850$ MeV, which possibly corresponds to the position of the
wide peak in $e^+e^- \to J/\psi D\bar D$ in \cite{Pakhlov}. This
enhancement of the shift is due to a much larger overlap matrix
element in the $^3P_0$ state. A similar situation was discussed in
\cite{Kalashnikova:2005ui}.

An interesting situation occurs in the case of the $1^{+-}$ state
with the $2~^1P_1$ pole at bare position 3959  MeV. Here the
coupling to the $DD^*$  channel is much weaker, than  in the case
of the $1^{++}$ state, and the main coupling is with the closed
$D^*D^*$ channel, which defines the destiny  of the bare pole.
When we renormalize the contribution of the $D^*D^*$ channel with
the coefficient $\kappa$, as discussed above, the final shift for
$\kappa= 0.5$ is around 45 MeV. In Fig. \ref{fig.1P1} we
demonstrate how the production cross section changes with
$\kappa$, and one can see, that the resulting width at
$\kappa=0.25 \div 0.5 $ is around $\Gamma\sim 50$ MeV.

In conclusion, we have calculated the amplitudes of CC processes
connecting $Q\bar Q$ and $(Q\bar q)(\bar Q q)$ systems via the
decay matrix element, Eq.(\ref{Eq.2}), involving realistic wave
functions of all hadrons involved, and  the CC constant
$\gamma(M_\omega)$, fitted earlier to bottomonium and charmonium
transitions. For the concrete case of the $1^{++}$ state of
charmonium we have found pole structure and production cross
section. At small CC two poles correspond to the complex
conjugated poles of one Breit-Winger resonance of the $2^3 P_1$
state of $Q\bar Q$ with the width $\Gamma\sim 35$ MeV. For
increasing CC one of the poles approaches the thresholds and
another moves away, as a result this  bare resonance flattens,
while a sharp cusp appears first at the $D_+D_-^*$ and then at the
$D_0D_0^*$ threshold at $\sim 3872$ MeV. This latter situation
with the sharp narrow cusp at the $D_0D_0^*$ threshold and absence
of any other structures (except for a tiny cusp at higher
threshold) including the region around 3940 MeV corresponds to the
observed production yield \cite{Pakhlova:2008di}. We conclude that
our dynamical mechanism explains properties (1)-(4), in
particular, why the resonance $X(3872)$ is at the lower, but not
the higher threshold, why it is so narrow, and why the original
$2^3P_1$ state of charmonium is not seen in experiment.

An alternative and close in spirit approach was developed recently
in \cite{Kalashnikova:2005ui,Kalashnikova:2009gt}. Our analysis
partly supports the conclusion in \cite{Zhang:2009bv}, that
``$X(3872)$ may be of ordinary $c\bar c$ $2^3P_1$ state origin".
Our results  differ from those of \cite{Ortega:2010qq}, where two
$1^{++}$ states were found, one associated with $X(3872)$, and
another with $X(3940)$. In a recent review \cite{Coito:2010cq} the
CC analysis of the $X(3872)$ and $X(3940)$ was reported with the
conclusion, that both states cannot be reproduced in the exploited
model simultaneously. This result is in common with ours, since in
our case the broad enhancement due to the second pole is near the
$D_0D_0^*$ threshold and cannot be associated with $X(3940)$.

The authors are grateful to Yu.S.Kalashnikova for numerous
discussions and useful advices, to A.M.Badalian for useful
comments, and to M.V.Danilov, G.Pakhlova and P.N.Pakhlov  for
discussions and suggestions. The financial support of Grant No.
09-02-00629a is gratefully acknowledged.

\end{document}